\documentclass[%
 reprint,
 superscriptaddress,
 footinbib,
 amsmath,amssymb,
 aps,
 prb,
floatfix,
]{revtex4-2}

\usepackage{graphicx}% Include figure files
\usepackage{dcolumn}% Align table columns on decimal point
\usepackage{bm}% bold math
\usepackage{xcolor}
\usepackage{soul}
\usepackage{circledsteps}
\newcommand{\CCircled}[2][]{\begingroup
\pgfkeys{/csteps/.cd,inner color=.,#1}%
\ifmmode
\Circled{$#2$}%
\else
\Circled{#2}%
\fi
\endgroup}

\newcommand{\kp}{$\mathbf{k\cdot p}$ }
\newcommand{\kvec}{\mathbf{k}}
\newcommand{\kvecT}{\mathbf{k}^\text{T}}
\newcommand{\zerovec}{\mathbf{0}}
\newcommand{\zerovecT}{\mathbf{0}^\text{T}}
\newcommand{\rperp}{\mathbf{r}_{\perp}}

\usepackage{physics}
\newcommand{\elu}{\ket{\frac{1}{2},\frac{1}{2}}_\text{EL}}
\newcommand{\eld}{\ket{\frac{1}{2},-\frac{1}{2}}_\text{EL}}
\newcommand{\hhu}{\ket{\frac{3}{2},\frac{3}{2}}_\text{HH}}
\newcommand{\hhd}{\ket{\frac{3}{2},-\frac{3}{2}}_\text{HH}}
\newcommand{\lhu}{\ket{\frac{3}{2},\frac{1}{2}}_\text{LH}}
\newcommand{\lhhd}{\ket{\frac{3}{2},-\frac{1}{2}}_\text{LH}}
\newcommand{\sou}{\ket{\frac{1}{2},\frac{1}{2}}_\text{SO}}
\newcommand{\sod}{\ket{\frac{1}{2},-\frac{1}{2}}_\text{SO}}

\newcommand{\cev}[1]{\reflectbox{\ensuremath{\vec{\reflectbox{\ensuremath{#1}}}}}}

\usepackage{color}
\usepackage{bbm}

\begin{document}

%\preprint{APS/123-QED}

\title{Band structure of $n$- and $p$-doped core-shell nanowires}% Force line breaks with \4\

\author{Andrea Vezzosi}
\email{andrea.vezzosi@unimore.it}
\affiliation{Dipartimento di Scienze Fisiche, Informatiche e Matematiche, Universit\`a di Modena e Reggio Emilia, Via Campi 213/a, 41125 Modena, Italy}
\affiliation{Centro S3, CNR-Istituto Nanoscienze, Via Campi 213/a, 41125 Modena, Italy}

\author{Andrea Bertoni}
\email{andrea.bertoni@nano.cnr.it}
\affiliation{Centro S3, CNR-Istituto Nanoscienze, Via Campi 213/a, 41125 Modena, Italy}

\author{Guido Goldoni}
\email{guido.goldoni@unimore.it}
\affiliation{Dipartimento di Scienze Fisiche, Informatiche e Matematiche, Universit\`a di Modena e Reggio Emilia, Via Campi 213/a, 41125 Modena, Italy}
\affiliation{Centro S3, CNR-Istituto Nanoscienze, Via Campi 213/a, 41125 Modena, Italy}

\date{\today}

\begin{abstract}
We investigate the electronic band structure of modulation-doped GaAs/AlGaAs core-shell nanowires for both $n$- and $p$-doping. We developed an 8-band Burt-Foreman \kp Hamiltonian approach to describe coupled conduction and valence bands in heterostructured nanowires of arbitrary composition, growth directions, and doping. Coulomb interactions with the electron/hole gas are taken into account within a mean-field self-consistent approach. We map the ensuing multi-band envelope function and Poisson equations to optimized, non-uniform real-space grids by the finite element method. Self-consistent charge density, single-particle subbands, density of states and absorption spectra are obtained at different doping regimes. For $n$-doped samples, the large restructuring of the electron gas for increasing doping results in the formation of quasi-1D electron channels at the core-shell interface. Strong heavy-hole/light-hole coupling of hole states leads to non parabolic dispersions with mass inversion, similarly to planar structures, which persist at large dopings, giving rise to direct LH and indirect HH gaps. In $p$-doped samples the hole gas forms an almost isotropic, ring-like cloud for a large range of doping. Here, as a result of the increasing localization, HH and LH states uncouple, and mass inversion takes place at a threshold density. A similar evolution is obtained at fixed doping as a function of temperature. We show that signatures of the evolution of the band structure can be singled out in the anisotropy of linearly polarized optical absorption.
\end{abstract}

\maketitle

%\tableofcontents

\section{\label{sec:Introduction}Introduction}

Among III–V compound semiconductor nanostructures, radially heterostructured nanowires represent an increasingly investigated, silicon-compatible perspective for applications in transistor-based electronic devices \cite{Jia2019} and opto-electronic devices \cite{Zhang2015,Zhang2016}. From the point of view of material quality, several issues have already been settled on the route to technological exploitation of nanowires or as a platform for coherent quantum phenomena. These include self-assisted growth \cite{Mandl,Breuer2011}, order and polytypism \cite{Krogstrup2010,Dick2011}, high-quality interfaces \cite{Zamani2018}, and multi-layer growth \cite{Royo2017}. One critical issue bridging between material science and device nanofabrication is the control of doping, for example in modulation doped heterostructures \cite{Spirkoska,Funk2013}  and radial $p$-$n$ junctions \cite{Goktas2018}. This is still a concern in terms of reproducibility between nanowires and homogeneity within each nanowire \cite{Kim2021,Jadczak2014}.

\begin{figure}[htp]
\includegraphics[scale=0.45]{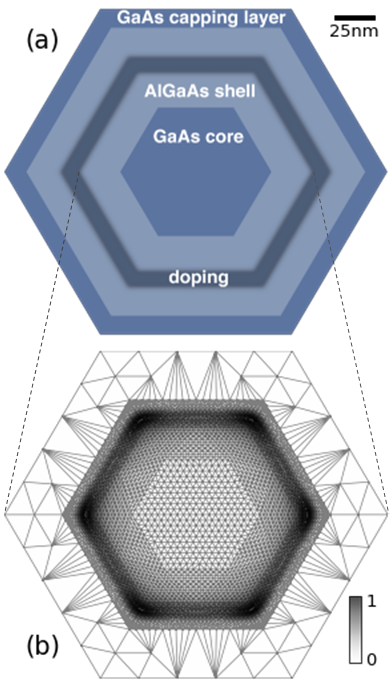}% Here is how to import EPS art
\caption{(a) Sketch of the section of the simulated CSNWs. The principal axes of the 2D-coordinate system are directed along the $[ 11\Bar{2} ]$ and  $[ \Bar{1}10 ]$ crystallographic directions. (b) An example of an optimized FEM grid used to solve the envelope function equation (\ref{eq:multibeq}) with superimposed self-consistent charge density (in grey scale arbitrary units). The grid stops at the doping layer, where the charge density is assumed to vanish. The grid used for the Poisson equation is different and extends to the outer boundary of the structure.
\label{fig:fig1}}
\end{figure}

As in the realm of planar heterostructures, GaAs-based nanomaterials play a special role also for nanowires. Ultra-high-mobility devices in planar GaAs/AlGaAs heterojunctions build on the modulation doping concept \cite{Dingle1978}, whereby dopants are incorporated in a higher gap AlGaAs layer, physically separated from the lower-gap layers, where carriers are confined, suppressing carrier-ionized impurity scattering. A corresponding, modulation doped radial heterostructure is schematically shown in Fig.~\ref{fig:fig1} \cite{Spirkoska}, which can be seen as a planar heterojunction with wrapped around layers. Carriers are confined in the GaAs core, while dopants are incorporated in an outer AlGaAs layer. Typically, a thin GaAs capping layer is included to prevent Al oxidation. While mobility is still improving in planar systems \cite{Chung2021}, where background impurities are the limiting factor, high mobility is  more difficult to achieve in core-multi-shell nanowires, though \cite{Funk2013,Balaghi2021}, and experimental and theoretical characterization is needed.

Due to comparable kinetic and Coulomb energies, in doped core-shell nanowires (CSNWs) electronic states \cite{Bertoni2011} and ensuing response functions \cite{Royo2014,Royo2015} are determined by the self-consistent field of free carriers, which, in turn depends on the concentration and type of doping \cite{Bertoni2011}, together with the Fermi level pinning at surface states \cite{Ishikawa1998,Alekseev2018a}. Hence, different doping regimes may result in distinct charge localization patterns \cite{Buscemi2016}. The ability to predict the band structure in doped CSNWs is therefore a complex task.  

Among the methods used, the envelope function approach stands out for its versatility and computational efficiency.
Single-band descriptions have been widely used, including non-perturbative electric and magnetic fields \cite{Bertoni2011,Royo2013,Royo2015a,Wu2019}. Multi-band \kp descriptions, which include spin-orbit coupling arising from valence states that are crucial to describe, e.g., optical properties \cite{Goldoni1996,Goldoni1997a}, have been employed for several classes of materials, taking into account composition modulations, crystallographic details and mesoscopic symmetries \cite{kishore2010electronic, kishore2012electronic, lassen2006electronic,pistol2008band, xu,luo2016band, ram2006wavefunction}. 
Spin-orbit coupling in the conduction band has been evaluated, also in presence of strong magnetic fields \cite{Wojcik2018,Wojcik2019,Wojcik2021}. However, a full description of the band structure of doped CSNWs in the different doping regimes including the self-consistent field arising from the free charge is still missing. 

In this paper we investigate the electronic band structure of modulation-doped GaAs/AlGaAs CSNWs with $n$- or $p$-type doping. We employ an 8-band Burt-Foreman \kp Hamiltonian approach, with Coulomb interactions with the electron/hole gas taken into account within a mean-field self-consistent approach. The numerical burden arising from the self-consistent solution of multi-band envelope function and Poisson equations is minimized by the use of the finite element method (FEM) with non-uniform real-space grids, optimized to different doping regimes. Self-consistent charge density, single-particle subbands, density of states and absorption spectra are then obtained. For strong $n$-doping, quasi-1D channel tend to form at the corners of the core-shell interface. Heavy-hole (HH)/light-hole (LH) couplings lead to non-parabolic dispersions with mass inversion in the valence band, similarly to planar structures, giving rise to direct LH and indirect HH gaps persisting at any doping density. In strongly $p$-doped samples, on the contrary, the hole gas forms an almost isotropic, ring-like cloud. As a result of the increasing localization, HH and LH states uncouple, and mass inversion takes place at a threshold density. Similar evolutions are obtained at fixed doping as a function of temperature. We suggest that signatures of the evolution of band structure can be traced in the anisotropy of linearly polarized optical absorption.

In Sec.~\ref{sec:Method} we outline our theoretical-computational methods, with detailed derivations reported in the Appendix. Emphasis is on the generality of the method, and mapping on optimized FEM grids. Band structures, density of states, projected charge densities and optical anisotropy are discussed in Sec.~\ref{sec:Result}, as a function of the doping density, separately for both $n$- and $p$-doped samples. 

\section{\label{sec:Method} Theoretical and computational methods}

\subsection{The \kp description}

To obtain the band structure of a modulation-doped CSNW, we have developed an 8-band \kp envelope function approach. Assuming translational invariance along the nanowire growth axis $z$, and the position vector $\mathbf{r}=(\rperp,z)$, the $n-$th eigenstate at the in-wire wave-vector $k_{z}$ are written as
\begin{eqnarray}
\Psi_n(\mathbf{r}, k_z) = \sum_{\nu=1}^{8} e^{i k_z z}  \psi_{n}^{\nu}(\mathbf{r}_{\perp},k_z) u^\nu(\mathbf{r})\,,
\label{eq:psi}
\end{eqnarray}
where $u^\nu(\mathbf{r})= \ket{J,J_z}$ is a Bloch basis function in the total angular momentum representation (see Eq.~\ref{eq:angular_momentum_basis}). We choose the quantization axis of $\mathbf{J}$ parallel to $z$. The coefficients $\psi^{\nu}_{n}(\mathbf{r}_{\perp},k_z)$ are the $\nu$-th component of the $n$-th solution of the multi-band envelope-function equation
\begin{multline}
\sum_{\nu=1}^{8} \big[ \hat{H}_{BF}^{\mu \nu}(\rperp, k_z)  - e V_{el}(\mathbf{r}_{\perp}) \delta^{\mu \nu} \big] \psi_{n}^{\nu}(\rperp, k_z) = \\  E_n(k_z) \psi_{n}^{\mu}(\rperp, k_z)\,,
\label{eq:multibeq}
\end{multline}
where $\hat{H}_{BF}$ is the $8 \cross 8$ Burt-Foreman Hamiltonian operator, with material-dependent parameters and including the band offsets, and $V_{el}$ represents the electrostatic potential generated by free carriers and fully ionized dopants.  

The operator $\hat{H}_{BF}$ is obtained from the \kp bulk Hamiltonian by replacing $k_x$ and $k_y$ with the corresponding differential operators. Material modulations are included by keeping track of the correct non-symmetrized operator ordering, as described in Appendix \ref{app:A}. This procedure yields a set of second order coupled partial differential equations which we numerically solve using FEM on an appropriate 2D grid \cite{ram2002finite, dhatt2012finite,zienkiewicz2005finite} with Dirichlet boundary conditions. The strongly non-parabolic subbands $E_n(k_z)$ and the corresponding envelope functions $\psi_{n}^{\nu}(\rperp, k_z)$ are finally determined on a uniform grid of wave vectors $k_z \in\left[ -k_{M}, k_{M} \right]$.

From the solutions of Eq.~(\ref{eq:multibeq}), we evaluate the total charge density
\begin{equation}
\rho(\rperp) = e \left[ n_h(\rperp) - n_e(\rperp) + n_D(\rperp) - n_A(\rperp) \right]
\label{eq:chdenstot}
\end{equation}
from the fully ionized donor or acceptor profiles, $n_D(\rperp)$ or $n_A(\rperp)$, respectively, and the free electron and hole charge densities given by, respectively, 

\begin{multline}
n_{e}(\rperp) = \sum_{n \in \mathrm{c.s.}} \sum_{\nu=1}^{8} \int_{-k_{M}}^{k_{M}} \frac{dk}{2 \pi} f(E_n(k),\mu, T)  \, \times \\ \times | \psi_{n}^{\nu}(\rperp, k)|^2,
\label{eq:chdensel}
\end{multline}
\begin{multline}
n_{h}(\rperp)  = \sum_{n \in \mathrm{v.s.}} \sum_{\nu=1}^{8} \int_{-k_{M}}^{k_{M}} \frac{dk}{2 \pi} (1-f(E_n(k),\mu, T)) \, \times \\ \times |\psi_{n}^{\nu}(\rperp, k)|^2\,,
\label{eq:chdensh}
\end{multline}
where the first summation runs over the conduction (valence) subband indices for electrons (holes). Here, $f(E,\mu, T)$ is the Fermi-Dirac distribution function, $\mu$ is the chemical potential, $T$ is the temperature and $k_B$ is the Boltzmann constant.  

In practice, Eq.~(\ref{eq:multibeq}) needs to be solved only in $\left[ 0, k_{M} \right]$, since eigenstates at negative wave vectors can be obtained (up to an arbitrary phase factor) applying the time reversal symmetry operator 
\begin{equation}
    \mathcal{T} = e^{-i \pi J_y} K \, ,
\label{eq:timorderop}    
\end{equation}
where $J_y$ is the $y-$component of the total angular momentum and $K$ is the complex conjugate operator.

The electrostatic potential $V_{el}(\rperp)$ is the solution of the Poisson equation with the source term given by the total charge density of the system, possibly with a material and position-dependent relative dielectric constant,
\begin{equation}
\grad \epsilon(\rperp) \grad V_{el}(\rperp) = - \frac{\rho(\rperp)}{\epsilon_0}\,.
\label{eq:poisson}     
\end{equation}
Again, Eq.~(\ref{eq:poisson}) is solved using FEM on a 2D grid with Dirichlet boundary conditions. The potential at the outer boundary of the CSNW is fixed to zero at the six edges of the outer layer of the structure. Note that the computational protocol allows to include arbitrary voltages at gates surrounding the nanowire \cite{Prete2021}, although we will not investigate this configuration here. 

The steps described above are iterated self-consistently until convergence in a seemingly Schr\"odinger-Poisson cycle, here generalized to a multi-band Hamiltonian. We stop iterations when the relative change of the charge density between two successive iterations falls below $10^{-3}$ at any node of the grid.

\begin{table}[b]
\caption{\label{tab:table1}%
 Material parameters used in the simulations at $T=20$ K. $E_{g}$ is the energy gap, $\Delta E_c$, $\Delta E_v$ are the  conduction and valence band offset values at the GaAs/Al$_{0.3}$Ga$_{0.7}$As interface, $\Delta_{so}$ is the split-off energy, $E_p$ is the bare Kane energy,  $E_{p}^{\text{rsc}}$ the rescaled Kane energy (Eq.~\ref{eq:ep}), $m_e$ is the effective conduction electron mass, $\gamma_{i}$ are the bare Luttinger parameters, $\tilde{\gamma}_{i}$ are the rescaled values (Eq.~\ref{eq:luparam}), $\epsilon_r$ is the relative dielectric constant and $a_{lc}$ is the lattice constant. The band structure parameters are taken from Ref.~\cite{band_parameters_reference} except for the band offset values. The latter have been determined assuming an offset ratio of $\Delta E_c:\Delta E_v = 63:37$, as recommended in Ref.~\cite{adachi2009properties}.}
\begin{ruledtabular}
\begin{tabular}{lccc}
\textrm{}&
\textrm{GaAs}&
\textrm{}&
\textrm{Al$_{0.3}$/Ga$_{0.7}$/As}\\
\colrule
$E_g$ [eV] & 1.518 &  & 1.936\\
$\Delta E_c$ [eV] & & 0.263 & \\
$\Delta E_v$ [eV] &  & 0.155 & \\
$\Delta_{so}$ [eV] & 0.341 &  & 0.323 \\
$E_{p}~/~E_{p}^{\text{rsc}}$ [eV] & 28.8~/~20.9 &  & 26.5~/~18.0 \\
$m_e$  & 0.067 &  & 0.092 \\
$\gamma_{1}~/~\tilde{\gamma}_{1}$  & 6.98~/~2.39 &  & 6.01~/~2.91 \\
$\gamma_{2}~/~\tilde{\gamma}_{2}$  & 2.06~/~-0.235 &  & 1.69~/~0.138 \\
$\gamma_{3}~/~\tilde{\gamma}_{3}$  & 2.93~/~0.635 &  & 2.48~/~0.928 \\
$\epsilon_{r}$  & 13.18 &  & 12.24 \\
$a_{lc}$ [nm]  & 0.56 &  &  \\
\end{tabular}
\end{ruledtabular}
\end{table}

To characterize bands states, a $k_z$-dependent spinorial analysis is useful. The contribution of any of the component of the envelope function can be estimated as 
\begin{equation}
    C_{n}^{\nu}(k_z) = \int  |\psi_{n}^{\nu}(\rperp, k_z)|^{2} \text{d}\rperp \, ,
    \label{eq:character_definition}
\end{equation}
with the normalization condition
\begin{equation}\nonumber
    \sum_{\nu=1}^{8}  C_{n}^{\nu}(k_z) = 1 \,,
\end{equation}
at each subband index $n$ and wave-vector $k_z$. When analysing electronic states, we shall classify states in terms of EL, HH, LH characters (see Appendix~\ref{app:A}, Eq.~(\ref{eq:angular_momentum_basis})), which are computed as
\begin{eqnarray}
\label{eq:character}
\begin{aligned}
C_\mathrm{EL}(k_z) & = & C_{n}^{1}(k_z) + C_{n}^{2}(k_z)\,, \\
C_\mathrm{HH}(k_z) & = & C_{n}^{3}(k_z) + C_{n}^{4}(k_z)\,, \\
C_\mathrm{LH}(k_z) & = & C_{n}^{5}(k_z) + C_{n}^{6}(k_z)\,. \\
\end{aligned}
\end{eqnarray}

We shall also plot  the projected probability distributions at $k_z=0$, defined as
\begin{eqnarray}
\label{eq:character2}
\begin{aligned}
\phi_\mathrm{EL}(\rperp) & = \sum_{\nu\in\{1,2\}} C^{\nu}_{n}(0)\frac{|\psi^{\nu}_{n}(\rperp,0)|^2}{\xi_{n}^{\nu}}\,, \\
\phi_\mathrm{HH}(\rperp) & = \sum_{\nu\in\{3,4\}} C^{\nu}_{n}(0)\frac{|\psi^{\nu}_{n}(\rperp,0)|^2}{\xi_{n}^{\nu}}\,, \\
\phi_\mathrm{LH}(\rperp) & = \sum_{\nu\in\{5,6\}} C^{\nu}_{n}(0)\frac{|\psi^{\nu}_{n}(\rperp,0)|^2}{\xi_{n}^{\nu}}\,,
\end{aligned}
\end{eqnarray}
where 
\begin{equation}\nonumber
    \xi_{n}^{\nu}=\max_{\rperp} |\psi^{\nu}_{n}(\rperp,0)|^2 \,.
\end{equation}

Additionally, we compute the projected density of states (PDOS) for any given component $\nu$ of the wave function,
\begin{equation}
    g_{\nu}(E) = \frac{1}{N} \sum_{n}^{subbands} \sum_{k}  C_{n}^{\nu}(k) \delta(E - E_{n}(k)) \,,
    \label{eq:pdos}
\end{equation}
where $N$ is the total number of points in $k$-space considered in the summation. Furthermore, for $n/p$-doped samples we evaluate the  self-consistent linear charge density of electrons/holes as
\begin{equation}
    \rho_{\text{lin}} = \int n_{e/h}(\rperp) \text{d}\rperp \,.
    \label{eq:linchdens}
\end{equation}

The calculation of the optical anisotropy proceeds as follows. In the dipole approximation, the interband absorption intensity of photons with energy $\hbar \omega$ and light polarization vector $\bm{\varepsilon}$ reads:

\begin{equation}
\label{eq:absorbtion}
\begin{aligned}
    &I_{\bm{\varepsilon}}(\hbar \omega) \propto \sum_{n \in \mathrm{v.s.}} \sum_{m \in \mathrm{c.s.}} \sum_{k} |M_{n\rightarrow m,k}^{\bm{\varepsilon}}|^{2} \, \times \\
    \times \, &[f(E_n(k))-f(E_m(k))] \, \delta[E_m(k)-E_n(k)+\hbar \omega] \, ,
\end{aligned}
\end{equation}
where $M_{n\rightarrow m,k}^{\bm{\varepsilon}}$ is the interband optical matrix element,\cite{bastard1990wave}
\begin{equation}
\label{eq:optmel}
\begin{aligned}
  M_{n\rightarrow m,k_z}^{\bm{\varepsilon}} \simeq \sum_{\mu \nu =1}^{8} &\matrixel{u^{\mu}}{\bm{\varepsilon} \cdot \mathbf{p}}{u^{\nu}} \times \\ \times &\int \text{d}\rperp \psi_{m}^{\mu^{*}}(\rperp,k_z) \psi_{n}^{\nu}(\rperp,k_z) \,.
\end{aligned}
\end{equation}
Note that doping, in addition to determine the envelope functions via the self-consistent field, enters Eq.~(\ref{eq:absorbtion}) through Fermi-Dirac distributions, which account for band filling effects when electron/hole subband edges approach the Fermi energy due to doping. For undoped structures the Fermi energy is well within the gap, and this term is almost equal to unity. In heavily doped structures, however, it inhibits interband transitions to the lowest subbands  which may be non-negligibly occupied. 

Finally, we compute the relative optical anisotropy $\beta$ between linearly polarized light along the wire axis, $I_{\varepsilon_z}$, and perpendicular to it along the $x$ direction, $I_{\varepsilon_x}$:
\begin{equation}
    \label{eq:optandef}
     \beta = \frac{I_{\varepsilon_{z}}-I_{\varepsilon_{x}}}{I_{\varepsilon_{z}}+I_{\varepsilon_{x}}} \,.
\end{equation}

\subsection{Numerical implementation details}

The above self-consistent 8-band \kp equations may result in a computationally intensive task, but a number of strategies can be implemented to keep the computational burden low and avoid the use of massively parallel architectures. Most of the strategies mentioned below take advantage of the flexibility of FEM which allows the use of non-uniform grids, which we generate by the \texttt{Free FEM} library \cite{MR3043640}. 

The \kp Hamiltonian is represented on a 2D hexagonal domain, partitioned in a $D_{6}$ symmetry-compliant, unstructured mesh of triangular elements. Since at different doping levels the charge density forms substantially dissimilar localization patterns \cite{Bertoni2011}, different density-optimized grids are used at different doping levels. A typical grid used for a high density regime is shown in Fig.~\ref{fig:fig1}(b), showing that the grid is denser where the charge density is localized.

We emphasize that the use of centro-symmetric grid is critical to correctly reproduce the expected degeneracies, without the need of extremely dense grids. Breaking the inversion symmetry of the grid would not only artificially split the orbital degeneracy expected in the conduction band (see Sec.~\ref{sec:intbs}), but also split the spin degeneracy, particularly in the strongly spin-orbit coupled valence band \cite{DeAndradaeSilva1997}. The need to maintain the inversion symmetry discourages the use of automatic adaptive grid methods. Hence, we use fixed, although optimized, non-uniform grids. 

In CSNWs which are at stage here, the charge density is confined to the GaAs core (although in a non trivial manner) and rapidly goes to zero inside the shell material; therefore, we use larger elements inside the shell with respect to the core and we require the envelope function to vanish somewhere inside the shell, typically at the doping layer. A typical grid used in the calculations is shown in Fig.~\ref{fig:fig1}(b). Finally, we found it convenient to use coarser grids during the self-consistent cycle, with optimized, finer grid used only in the last iterations.

For the 8-band \kp model the bound states of interest around the gap correspond to interior eigenvalues of the Hamiltonian matrix. To compute the charge density via Eqs.~(\ref{eq:chdensel}), (\ref{eq:chdensh}) the sum is restricted to few tens of subbands (usually $n_{max}$=60 for the electrons and $n_{max}$=100 for holes), and iterative methods are preferable. We use the Arnoldi method \cite{sorensen1997implicitly}, implemented in the \texttt{ARPACK} library \cite{lehoucq1998arpack}, together with the shift-and-invert approach, where the original problem is recast to target the largest eigenvalues. This approach provides faster convergence and enables the search for $n_{max}$ eigenvalues around an energy value $E_{search}$. Thus, since for both $n$- and $p$-doping the occupation of the minority carrier is negligible, during the self-consistent cycle one needs to solve only for the conduction or the valence band structure, respectively, by properly choosing $E_{search}$ \footnote{Note that the convergence of the diagonalization is affected by the actual value of $E_{search}$ \cite{andlauer2009optoelectronic}. For this reason, computational times can be further reduced by using the minimum (maximum) conduction (valence) band eigenvalue at  $E_{n}^{min}(k_z)$ ($E_{n}^{max}(k_z)$) as $E_{search}$ for the next diagonalization at $k_{z}+\Delta k_{z}$.}. The full band structure is then calculated only in the final converged self-consistent potential.

The Poisson equation is solved on a single specific mesh extending over the entire 2D domain of the heterostructure. To go back-and-forth between the grids of the envelope function and Poisson solver, as well as between different grids used during the self-consistent cycle, we make use of 2D linear interpolation. 
To achieve the convergence of the self-consistent protocol we rely on the modified second Broyden's method \cite{EYERT1996271, Johnson1988, Vanderbilt1984, broyden1965class} when updating the electrostatic potential at the current iteration. The inverse Jacobian is updated using the information from $M=8$ previous iterations. We fixed the weight corresponding to the first iteration to $w_{0}=0.01$, while all the other weights $w_{m}$, with $m=1,...,M-1$, are computed as suggested in Ref.~\cite{Johnson1988}.
The simple mixing parameter $\alpha$ is fixed to $0.05$.

Before the simulation starts, the mesh is processed through a bandwidth reduction procedure leveraging the reverse Cuthill–McKee algorithm \cite{cuthill1969reducing} implemented within the \texttt{SciPy} library \cite{2020SciPy-NMeth}. This is done in order to obtain tightly banded sparse matrices from the FEM discretization.

The above self-consistent numerical protocol and ancillary calculations have been implemented in a \texttt{python} library. A typical run uses a grid of about 7000 triangular elements and 3500 nodes for the \kp problem and 10-20 self-consistent iterations. A run on a single node architecture equipped with 16 2.60 GHz Intel Xeon E5-2670 processor cores takes about 6 hours CPU time.

\section{\label{sec:Result}Results}

We simulate a typical modulation-doped structure~\cite{Funk2013} consisting of a GaAs hexagonal core with an edge-to-edge distance of 80 nm surrounded by a 50-nm-wide Al$_{0.3}$Ga$_{0.7}$As shell and a GaAs capping layer of thickness 10 nm [see Fig.~\ref{fig:fig1}(a)]. The 2D-coordinate system has the $x$- and $y$-axes directed along the $[ 11\Bar{2} ]$ and $[ \Bar{1}10 ]$ crystallographic directions respectively. Buried inside the shell, at a distance of 20 nm from the core-shell interface, a 10-nm-thick layer is doped at a constant density $n_{D}$ of donors or $n_{A}$ of acceptors. All calculations discussed below are performed at $T=20$ K, except in Sec.~\ref{sec:tdependence}. The chemical potential $\mu$ is fixed at the mid-gap value of GaAs \cite{Alekseev2018a}. 

\subsection{\label{sec:intbs} Band structure of the undoped material}

\begin{figure}[htp]
\includegraphics[scale=0.7]{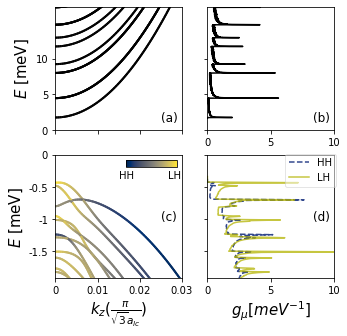}
\caption{\label{fig:intbs} Conduction (a) and valence (c) subbands of an undoped GaAs/AlGaAs CSNW (see text for parameters). In (c) the hue/color represents the spinorial character in terms of HH and LH, according to Eqs.~(\ref{eq:character}). Conduction (b) and valence (d) PDOS for different spinor components. The zero of the energy in each panel is taken at the bulk band edge of GaAs for conduction and valence band, respectively.}
\end{figure}

\begin{figure}[htp]
\includegraphics[scale=0.5]{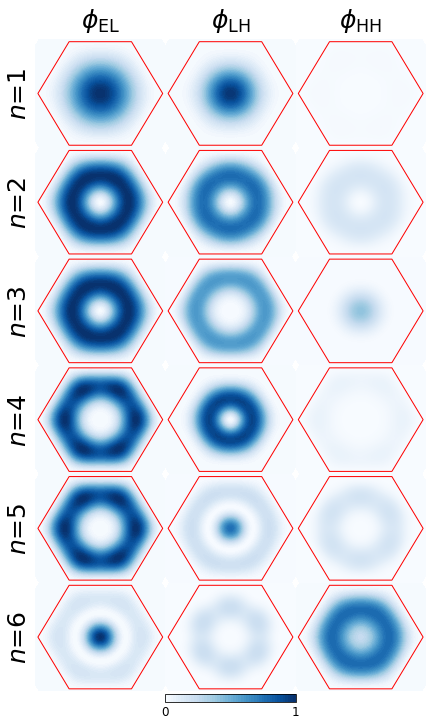}
\caption{\label{fig:envbs_kz0} Projected probability distributions [Eqs.~(\ref{eq:character2})] of the six lowest conduction band (1st column) and six highest valence band states (2nd and 3rd column) at $\Gamma$ for the undoped material of Fig.~\ref{fig:intbs}.}
\end{figure}

As a reference for calculations of the band structure of doped CSNWs to be discussed in the next sections, we first consider an undoped sample and analyze the conduction and valence bands subbands, which are shown in Fig.~\ref{fig:intbs}(left), together with the corresponding PDOSs (right). These are best analyzed together with the projected probability distributions of the EL, HH, and LH spinor components [see Eqs.~(\ref{eq:character}), (\ref{eq:character2})] at $k_z=0$, which are shown separately in Fig.~\ref{fig:envbs_kz0} \footnote{The SO component for these states is negligible and it is not shown here}.

We first consider conduction states. Due to the large gap of GaAs, which disentangles conduction and valence bands in the \kp Hamiltonian, conduction subbands [Fig.~\ref{fig:intbs}(a)] show an almost pure EL character with parabolic dispersion and ensuing $1/\sqrt{\mathrm{energy}}$ PDOS [Fig.~\ref{fig:intbs}(b)]. In a system with $D_{6h}$ symmetry, assuming a perfectly isotropic band structure, we expect the ground state to be non degenerate, while the second/third and fourth/fifth doublets are degenerate \cite{Ferrari2009}. Here, anisotropic residual interactions with the valence band remove the degeneracies by $\sim 10^{-3} \text{meV}$, a quantity which cannot be distinguished in Fig.~\ref{fig:intbs}. Indeed, the single/double degeneracy of the levels is easily recognized in the height of the peaks of the PDOS. 

As shown in Fig.~\ref{fig:envbs_kz0}(left column), the lowest conduction state is $1s$-like in the center, while the nearly degenerate doublets are ring-like states with an increasing modulation in the corners of the hexagonal confining potential. The 6-th state is again a non-degenerate state with a $2s$ character. Higher levels (not shown here) have maxima on the corners of the hexagon and nodes along the facets or vice versa.\cite{Ferrari2009, kokurin2020electronic}

The valence subbands [Fig.~\ref{fig:intbs}(c)] are of course denser in energy than conduction subbands, due to the larger mass of holes. The LH-HH mixing, which is small but finite also at $\Gamma$, leads to a strongly non-parabolic dispersion of the subbands with $k_z$. As shown by the color code of the lines, the two highest subbands have a predominant LH character at $\Gamma$, which is also shown by the corresponding distribution functions in Fig.~\ref{fig:envbs_kz0}(center and right columns) \footnote{Note that the quantization axis of the total angular momentum $\mathbf{J}$ is chosen along the free direction, i.e., the nanowire axis $z$, not along the confinement direction, as usually done in quantum wells; hence, the HH/LH labelling  of the highest valence state is opposite to the quantum well case, see, e.g., ~\protect\cite{Los1996}}. In between several subbands in Fig.~\ref{fig:intbs}(c) pointing downward and with a strong LH character at $\Gamma$, we recognize a mixed character state (the 3rd subband) and an almost HH subband (the 6-th state) (see also Fig.~\ref{fig:envbs_kz0}). These two subbands strongly couple at finite wave vectors (note from the hue that these two subbands exchange their HH-LH character), causing a strong camel's back dispersion of the third subband and a corresponding peak in the PDOS at $\sim -0.54 \text{meV}$ with ~50\% character of either HH and LH components. All in all, the LH character dominates the PDOS, which agrees with Ref.~\onlinecite{xu}. Note that band crossings of the third subband can be traced to states belonging to different irreducible representations of the $C_{3v}$ double symmetry group of [111] oriented nanowires with hexagonal cross-section \cite{liao2016electronic,Liao_2016,Cygorek2020}.

The probability distributions of HH and LH states shown in Fig.~\ref{fig:envbs_kz0} are either $s$-like or ring-like (arising from a quadrupolar symmetry of the real/imaginary parts of the envelope functions), similarly to corresponding conduction states, although of course the ordering is different, as HH- and LH-like states interlace. No orbital degeneracies are expected, since the strongly anisotropic bulk valence band structure does not share the hexagonal symmetry of the confinement.

We finally note that all electronic states are doubly spin-degenerate, due to the centro-symmetric symmetry of the system (which is carefully preserved by the FEM grid), which will hold true in all calculations throughout \footnote{It should be noted that our \kp model has an higher symmetry with respect to an atomistic Hamiltonian due to lack of bulk inversion symmetry (BIA) in III-V materials. This would remove the double degeneracy of our \kp subbands with tiny spin-splittings in the micro-eV range \cite{Liao_2016}.}.

\begin{figure}[htp]
\includegraphics[scale=0.55]{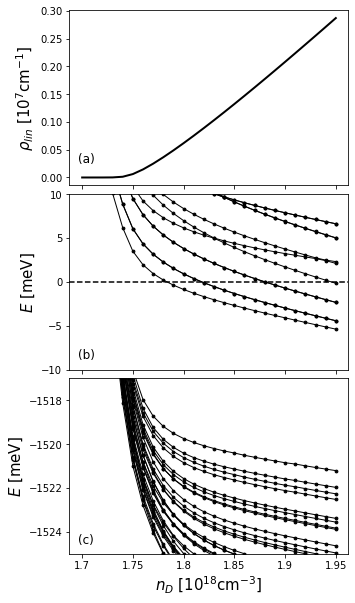}
\caption{\label{fig:dopN_levels_chlin} (a) Linear free charge density [Eq.~(\ref{eq:linchdens})], (b) conduction subband energies at $k_z = 0$, and (c) valence subband energies at $k_z = 0$ as a function of the doping density $n_{D}$. Energies are referred to the Fermi level.}
\end{figure}

\begin{figure*} [htp]
\includegraphics[scale=0.55]{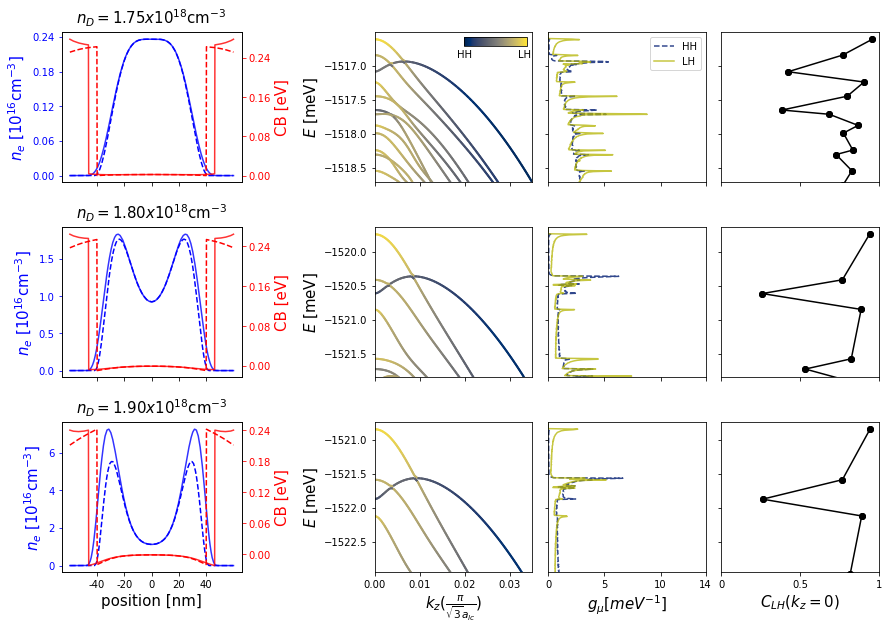}
\caption{\label{fig:dopN_vale} Left column: free charge density distribution $n_e$ (blue) and self-consistent conduction-band profile CB (red) shown along the edge-to-edge (dashed line) and corner-to-corner (full line) directions of the CSNW section for $T=20$ K at selected values of $n_{D}$, as indicated. Doping increases from top to bottom. Middle left and middle right columns: valence subbands and PDOS, respectively, corresponding to the doping density and self-consistent potential of the left panels. 
%\st{Hue/color as in} Fig.~\ref{fig:intbs}. 
The hue/color represents the spinorial character in terms of HH and LH, according to Eqs.~(\ref{eq:character}). Right column: LH-character of each subband at $\Gamma$.}
\end{figure*}

\begin{figure*}[htp]
\includegraphics[scale=0.25]{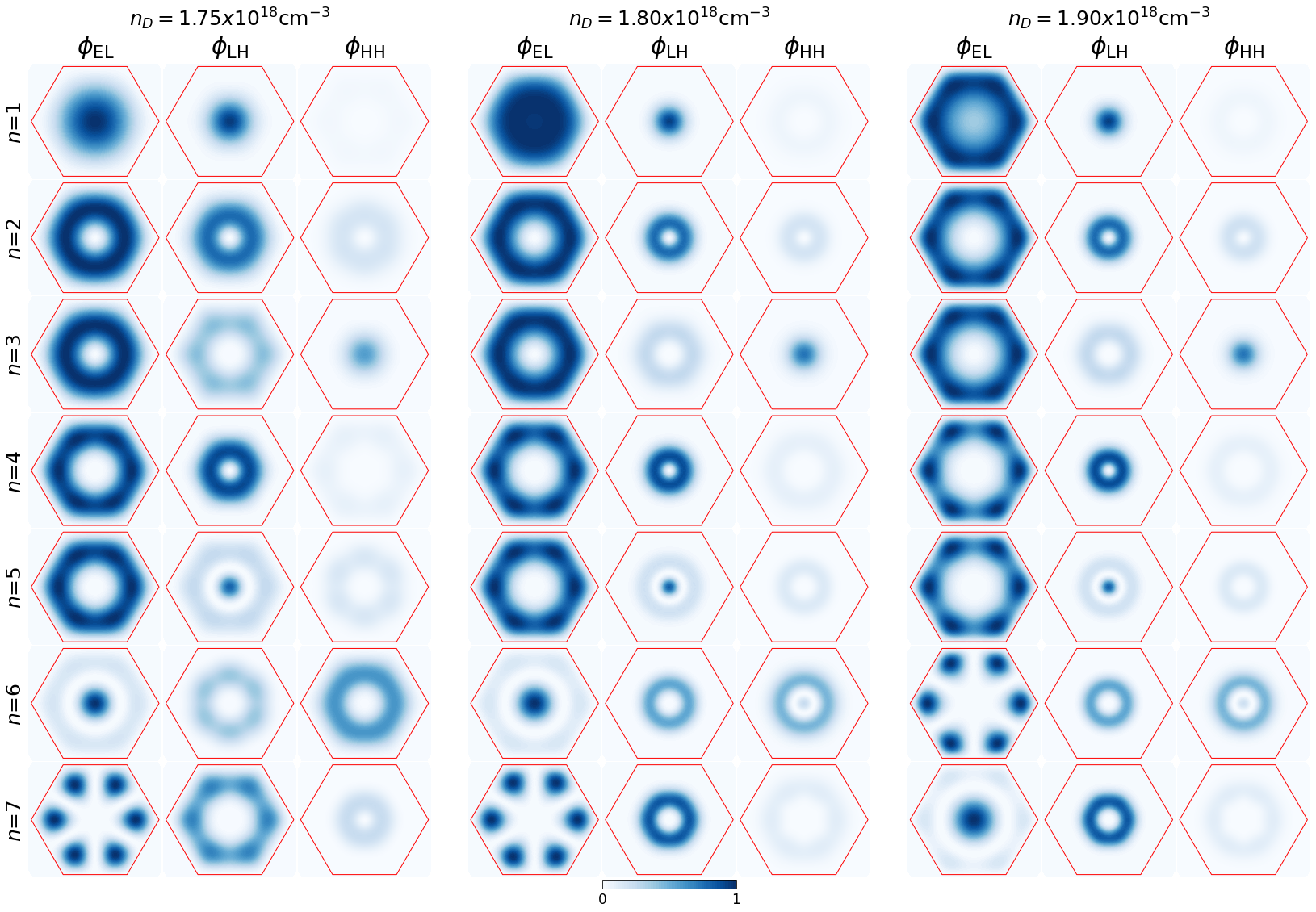}
\caption{\label{fig:dopN_env_kz0} Projected probability distributions [Eqs.~(\ref{eq:character2})] for the seven lowest conduction and seven highest valence subbands at the same selected doping densities of Fig.~\ref{fig:dopN_vale}, as indicated. Each column corresponds to the EL, LH, HH component, as indicated.}
\end{figure*}

\subsection{\label{sec:ndopbs}$n$-doping}

We now consider $n$-doped samples with increasing doping density $n_{D}$, up to high-doping regimes. As shown in Fig.~\ref{fig:dopN_levels_chlin}(a), the self-consistent linear charge density [Eq.~(\ref{eq:linchdens})] increases almost linearly for large doping, while an increasing number of conduction subbands fall below the Fermi energy [Fig.~\ref{fig:dopN_levels_chlin}(b)]. The evolution of the (unoccupied) valence band states at $\Gamma$ is also shown in Fig.~\ref{fig:dopN_levels_chlin}(c). 

The evolution of the localization of the self-consistent charge density and the corresponding electrostatic potential, shown in Fig.~\ref{fig:dopN_vale}(left), is not trivial. With increasing doping, the charge density evolves from a small, isotropic charge distribution in the core of the structure to a larger, ring-like charge density distribution, and finally to a charge density which is primarily located in the corners of the core, as can be inferred by comparing the edge-to-edge and corner-to-corner profiles in Fig.~\ref{fig:dopN_vale}. This is in agreement with single-band self-consistent calculations ~\cite{Bertoni2011,Sitek2019}, as expected from the nearly pure EL character of conduction subbands.  

Conduction subbands retain a trivial parabolic dispersion regardless of the doping level (and type), which is therefore not shown here. However, it is still interesting to consider the evolution of the localization of the conduction envelope functions shown in terms of the projected density distribution $\phi_{\mathrm{EL}}(\rperp)$ in Fig.~\ref{fig:dopN_env_kz0} (left columns in each panel), with increasing doping (panels from left to right). For each of the seven lowest levels, the larger the doping, the more localized is $\phi_{\mathrm{EL}}(\rperp)$ at the core-shell interface. For the largest doping shown here, all subbands feature a clear six-fold symmetry induced by the heterostructure confining potential. Note that the ordering of the levels in terms of symmetry depends on the level of doping, as seen from the ``exchange'' of the 6-th and 7-th levels with increasing doping.

Although for $n$-doping the charge density is determined by conduction band states, the valence band structure does have an evolution as well, due to the restructuring of the free charge density and ensuing change in the self-consistent confining electrostatic potential shown in Fig.~\ref{fig:dopN_vale}. The valence band structure shown in Fig.~\ref{fig:dopN_vale} (second column) shows a downward shift of the subbands and an increase of the inter-subband gaps, due to the increased localization energy at the core-shell interface. The $k_z=0$ character  (Fig.~\ref{fig:dopN_vale}, right column) at low doping is $\sim\!10 \divisionsymbol 30$\% LH for most states, except for the ground level which is almost completely LH, and  two states which stand out with a strong HH character. Increasing doping increases the gaps, but does not change much the subband dispersions. At the largest doping shown here, the PDOS is dominated by i) a LH peak near the gap, and ii) two overlapping peaks, one arising from a LH band and one from a HH band. Note, however, that the latter HH peak arises from the camel's back subband with a maximum at a finite $k_z$ and, therefore, an indirect gap with the conduction band. 

Figure~\ref{fig:dopN_env_kz0} shows that as doping increases holes tend to be more localized in the center, with a mostly isotropic distribution. This is at difference with conduction states which move towards the GaAs/AlGaAs interface at larger doping densities, and it is due to the opposite sign of the electrostatic energy. Note that, as already noted for EL states, also for HH and LH states the order in terms of symmetry is not preserved as doping is swept. For example the 7-th level changes both character and orbital symmetry as $n_D$ moves from 1.75 to 1.80 $10^{18}$\,cm$^{-3}$.

\subsection{\label{sec:Pdopbs}$p$-doping}

\begin{figure}[htp]
\includegraphics[scale=0.55]{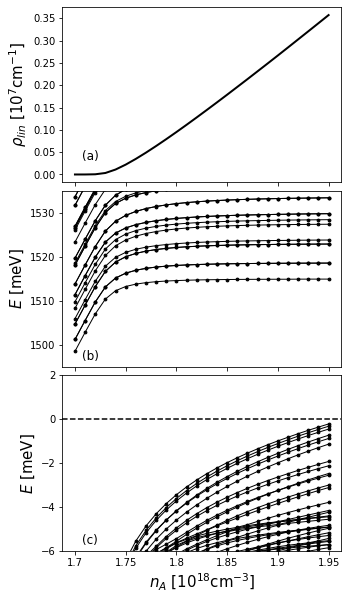}
\caption{\label{fig:dopP_levels_chlin} 
%Same as Fig.~\ref{fig:dopN_levels_chlin} for $p$-doped samples with doping density $n_A$.
(a) Linear free charge density [Eq.~(\ref{eq:linchdens})], (b) conduction subband energies at $k_z = 0$, and (c) valence subband energies at $k_z = 0$ as a function of the doping density $n_{A}$. Energies are referred to the Fermi level.}
\end{figure}

\begin{figure*}[htp]
\includegraphics[scale=0.55]{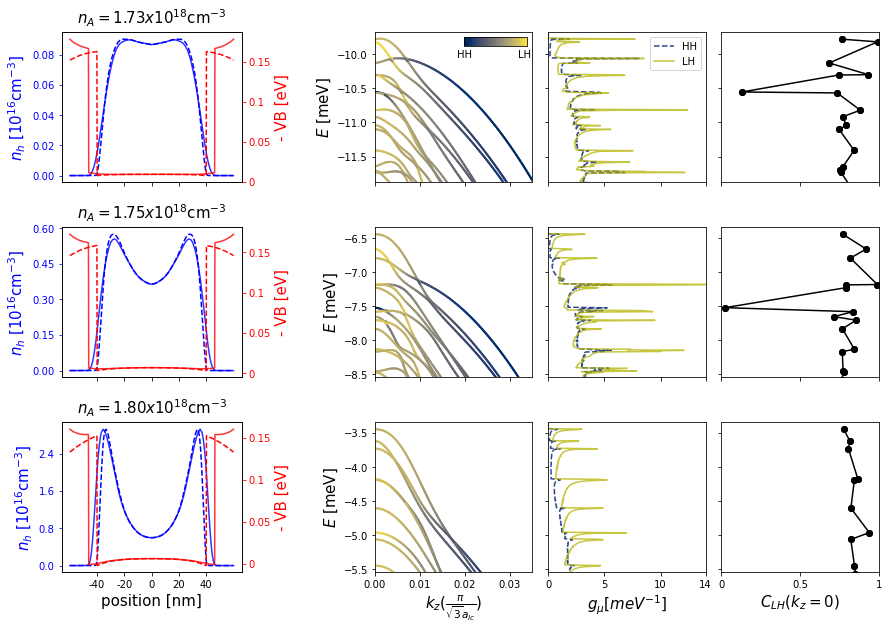}
\caption{\label{fig:dopP_vale} %Same as Fig.~\ref{fig:dopN_vale} for $p$-doped samples with doping density $n_A$.
Left column: free charge density distribution $n_h$ (blue) and self-consistent valence-band profile VB (red) shown along the edge-to-edge (dashed line) and corner-to-corner (full line) directions of the CSNW section for $T=20$ K at selected values of $n_{A}$, as indicated. Doping increases from top to bottom. Middle left and middle right columns: valence subbands and PDOS, respectively, corresponding to the doping density and self-consistent potential of the left panels. The hue/color represents the spinorial character in terms of HH and LH, according to Eqs.~(\ref{eq:character}). Right column: LH-character of each subband at $\Gamma$.}
\end{figure*}

\begin{figure*}[htp]
\includegraphics[scale=0.25]{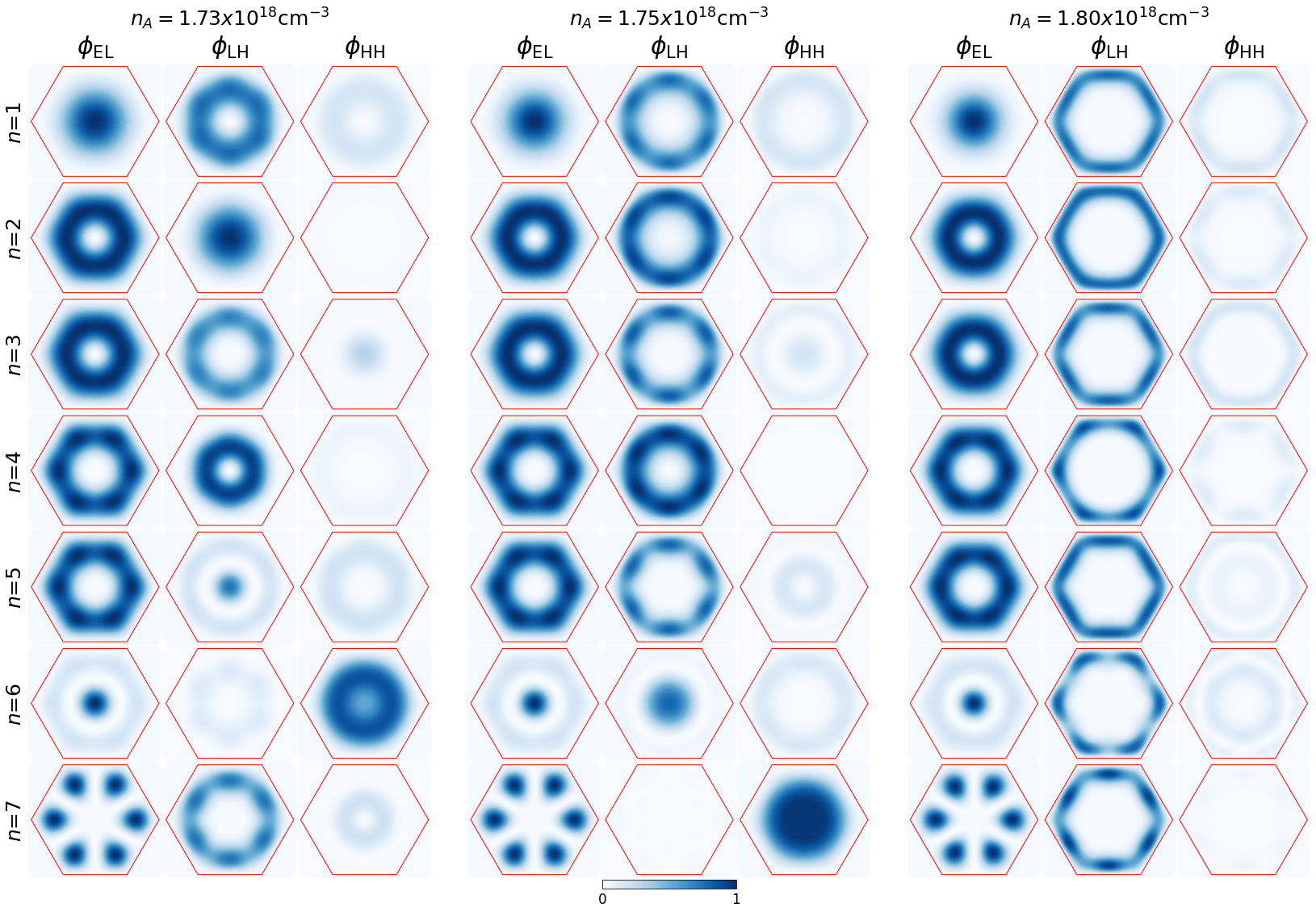}
\caption{\label{fig:dopP_env_kz0} %Same as in Fig.~\ref{fig:dopP_env_kz0} for $p$-doped samples. 
Projected probability distributions [Eqs.~(\ref{eq:character2})] for the seven lowest conduction and seven highest valence subbands at the same selected doping densities of Fig.~\ref{fig:dopP_vale}, as indicated. Each column corresponds to the EL, LH, HH component, as indicated.}
\end{figure*}

We next discuss the results for \textit{p}-doped materials, focusing on the effects of an increasing acceptor density $n_{A}$ on the band structure and the hole charge density localization.

Figure \ref{fig:dopP_levels_chlin} shows a linear increase of the free charge density after a threshold density of dopants. Note that the range of densities is similar with respect to the $n$-doping case, despite the very different parameters and, as we shall see below, charge localization.

Indeed, the free charge $n_h$, shown in  Fig.~\ref{fig:dopP_vale}(left) at selected values of the acceptor density $n_A$, shows a dip in the center already at weak doping, which is consistent with the larger mass and lower confinement energy of holes with respect to conduction electrons. As the acceptor doping density $n_A$ increases, the charge progressively moves toward the interfaces to minimize Coulomb energy, in analogy with the $n$-doping case, but at difference with the latter case the hole gas remains remarkably isotropic, a seen by comparing the edge-to-edge and corner-to-corner profiles which nearly coincide in Fig.~\ref{fig:dopP_vale}(left). In other words, the hole charge forms a uniform gas with a cylindrical shape and little resemblance to the host hexagonal confining potential up to these doping densities.

As $n_A$ is swept, the conduction levels [Fig.~\ref{fig:dopP_levels_chlin}(b)] shift in energy with respect to the Fermi level and finally stabilize, while an increasing number of hole subbands approach the Fermi energy and contribute to the free charge. Note that at large dopings, hole levels separate in a low-energy and a high-energy branch, which correspond to increasingly LH- and HH-like levels, respectively. 

At difference with the $n$-doping case, the hole band structure is strongly affected by $p$-doping, as exemplified in Fig.~\ref{fig:dopP_vale}. This is due to the different localization energies of HHs and LHs in the increasingly localizing self-consistent potential. A prominent effect can be seen by comparing Figs.~\ref{fig:dopP_vale} and \ref{fig:dopP_env_kz0}. The only strongly HH level (the 6-th level in Fig.~\ref{fig:dopP_env_kz0}, left panel) moves to lower energy due to the light mass. As a result, HH-LH mixing and related anticrossings are removed, the mass of the camel's back subband changes sign, and all bands point downward with a small mass at the large densities. 
Note that the PDOS at large doping is dominated by far by LH states. Furthermore, as a consequence of the reduced \kp coupling in the valence band at high doping densities, the hole energy levels at $\Gamma$ tend to group in 6-fold clusters [see Fig.~\ref{fig:dopP_levels_chlin}(c)] separated by gaps that increase with increasing $n_{A}$ \cite{Ferrari2009}.

Figure \ref{fig:dopP_env_kz0} shows that all highest valence subbands become strongly localized at the interfaces at high doping. Contrary to conduction electrons, however, which always tend to localize at the six corners, holes alternate subbands localized at the corners and at the facets, which is again in agreement with single-band calculations in Ref.~\cite{Bertoni2011}. Since the charge density is a convolution of these levels, the isotropy of the hole cloud noted above is justified. 

\begin{figure*}[htp]
\includegraphics[scale=0.55]{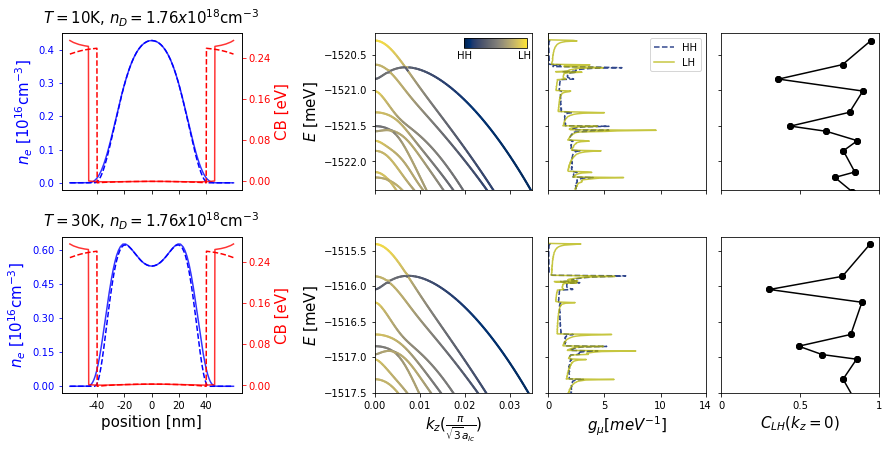}
\caption{\label{fig:dopN_valeT} Left column: free charge density distribution $n_e$ (blue) and self-consistent conduction-band profile CB (red) shown along the edge-to-edge (dashed line) and corner-to-corner (full line) directions of the CSNW section for $T=10$ K (top row) and $T=30$ K (bottom row) at the single doping density $n_{D} = 1.76 \cross 10^{18}\,\mbox{cm}^{-3}$. Middle left and middle right columns: valence subbands and PDOS, respectively, corresponding to the doping density, temperature and self-consistent potential of the left panels. The hue/color represents the spinorial character in terms of HH and LH, according to Eqs.~(\ref{eq:character}). Right column: LH-character of each subband at $\Gamma$.}
\end{figure*}

\begin{figure*}[htp]
\includegraphics[scale=0.55]{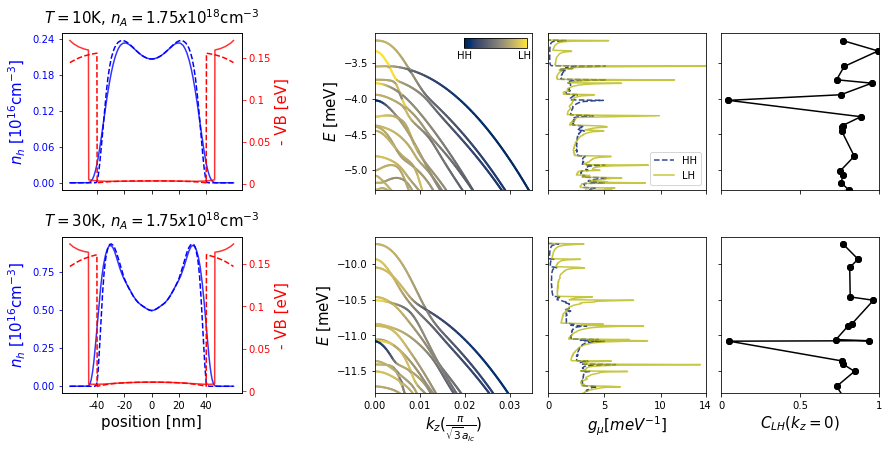}
\caption{\label{fig:dopP_valeT} Left column: free charge density distribution $n_h$ (blue) and self-consistent valence-band profile VB (red) shown along the edge-to-edge (dashed line) and corner-to-corner (full line) directions of the CSNW section for $T=10$ K (top row) and $T=30$ K (bottom row) at the single doping density $n_{A} = 1.75 \cross 10^{18}\,\mbox{cm}^{-3}$. Middle left and middle right columns: valence subbands and PDOS, respectively, corresponding to the doping density, temperature and self-consistent potential of the left panels. The hue/color represents the spinorial character in terms of HH and LH, according to Eqs.~(\ref{eq:character}). Right column: LH-character of each subband at $\Gamma$.}
\end{figure*}

We also note that, as doping is increased, there is no definite order of LH- and HH-like levels in term of symmetry/localization, due to the increasing hole confinement energy towards the core-shell interface which is different for HH and LH components. 

Finally, we note that, similarly to $n$-doped samples, minority carriers localize in the opposite direction, due to the opposite sign of the self-consistent potential. However, conduction electron are much more rigid and stable due to the light mass, hence showing little evolution with doping density, and in particular no symmetry inversion takes place. 

\subsection{Temperature dependence \label{sec:tdependence}}

The electronic states discussed above are the result of the competition between comparable energy scales in the meV range. As temperatures of $\sim 10$\,K are in the same energy range, we expect that small changes in temperature at this scale may bring about strong restructuring of the electronic system. As we shall see below, in general the effect of a temperature variation on the free-carrier charge density and the valence band structure are qualitatively analogous to the effects of a varying doping density.

In Fig.~\ref{fig:dopN_valeT} we consider an $n$-doped sample with donor density $n_{D}=1.76 \cross 10^{18}\,\mbox{cm}^{-3}$ at $T=10$\,K (top row) and $T=30$\,K (bottom row), which are above and below the temperature used in Sec.~\ref{sec:ndopbs}. Such temperature variations respectively increase or decrease the bulk-band gap values of 1 meV with respect to the values in Tab.~\ref{tab:table1} for both the core and the shell materials. As a result, the band-offset are unchanged, while the band structure parameters that are affected by a rescaling procedure are slightly modified. Starting at the lower temperature, the electronic charge density (left column) evolves from an isotropic charge density centered in the core to a ring-like density. This is similar to the effect of increasing doping, as in Fig.~\ref{fig:dopN_vale}, as the occupation probability of the levels above the chemical potential increases with temperature, and more charge populates the nanowire. Consistently with Fig.~\ref{fig:dopN_vale}, the valence band structure and PDOS are little affected by temperature in this range. However, the subbands are shifted in the opposite direction with respect to Fig.~\ref{fig:dopN_vale}. 

In Fig.~\ref{fig:dopP_valeT} we consider a $p$-doped sample with acceptor density $n_{A}=1.75 \cross 10^{18} \text{cm}^{-3}$  at the same two temperature as above. Again, increasing the temperature results in a greater hole charge density and a more pronounced charge depletion in the center due to the Coulomb interaction. Clearly, valence band states are more sensitive to changes in the charge density for $p$-doping. Indeed, Fig.~\ref{fig:dopP_valeT} shows that as temperature is increased, HH-like states move to lower energies, while HH-like subbands change their curvature downward. As a consequence, the PDOS undergoes a substantial restructuring, as all main features are LH-like. Note that in contrast to the case of a doping density variation, the valence band structure is shifted downward when the temperature increases.

\subsection{Optical anisotropy \label{sec:optan}}

Optical absorption in quasi-one-dimensional systems is dominated by excitonic and polarization effects induced by Coulomb interactions, not included in Eqs.~(\ref{eq:absorbtion}),(\ref{eq:optmel}) \cite{Rossi1996,Rossi1999}. However, the optical anisotropy between linearly polarized light along and transverse the nanowire axis, should be less sensitive to Coulomb effects \cite{Goldoni1996,Goldoni1997a}. On the other hand, while $x$-polarized light couples to HH states [see Eq.~(\ref{eq:angular_momentum_basis})], $z$-polarized light does not. Hence, $\beta$ is a sensitive probe of the orbital composition of valence band states \cite{Goldoni1996}.

\begin{figure*}[htp]
\includegraphics[scale=0.55]{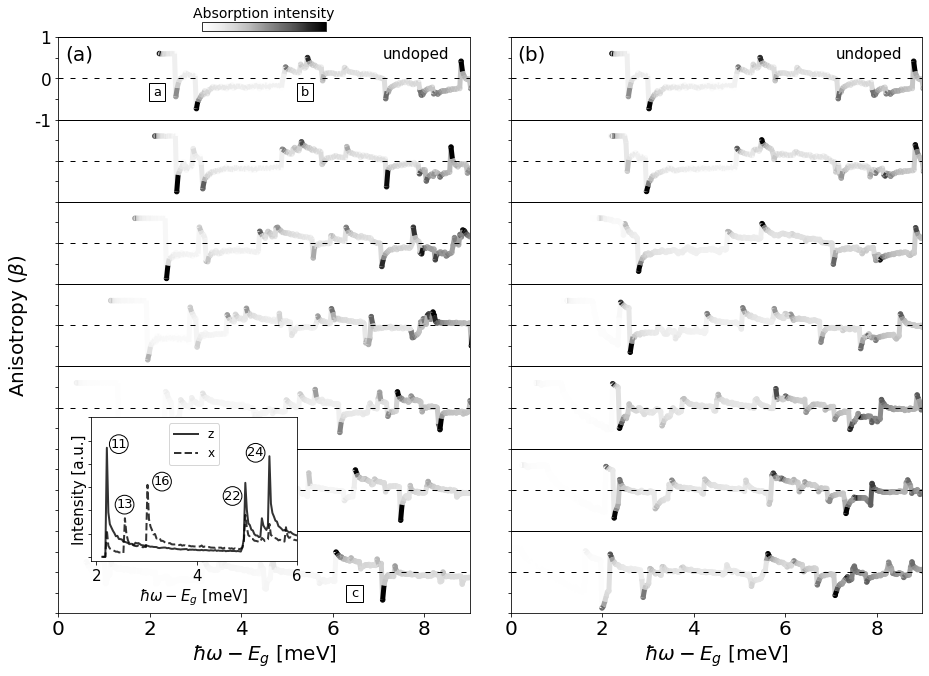}

\caption{\label{fig:opt} (a) Optical anisotropy $\beta$ [Eq.~(\ref{eq:optandef})] for $n$-doped samples at different donor concentrations. From top to bottom: undoped, 1.75, 1.77, 1.79, 1.82, 1.87, 1.90 $\cross10^{18}\,\text{cm}^{-3}$. Horizontal dashed lines indicate the zero reference, each panel extends vertically from -1 to +1. The gray hue represents the intensity of the corresponding absorption spectrum [Eq.~(\ref{eq:absorbtion})] at the given photon energy. $E_g$=1.518 eV is the band gap of GaAs at $T=20$ K. Inset: calculated absorption spectra of the undoped structure for linearly polarized light. Peaks are labelled with \CCircled[outer color=black]{\textit{mn}}, where $m$ is the index of the final conduction subband and $n$ the index of the initial valence subband involved in the optical transition. (b) Same as panel (a) but for $p$-doped samples. From top to bottom: undoped, 1.73, 1.75, 1.77, 1.79, 1.80, 1.82 $\cross10^{18}\,\text{cm}^{-3}$.}
\label{fig:optan}
\end{figure*}

In Fig.~\ref{fig:opt} we show the calculated relative optical anisotropy $\beta$ [Eq.~(\ref{eq:optandef})] at selected doping concentrations for $n$- (left) and $p$-doped (right) samples, respectively. Doping concentration increases from top to bottom in both panels. To emphasize the anisotropy of the more intense absorption peaks, the line darkness is modulated with the intensity of the absorption spectrum at the given photon energy. For reference, we also show in the inset single-particle absorption spectra in the two polarizations (for the undoped sample) with optical transitions from the $n$-th valence state to the $m$-th conduction state labelled \CCircled[outer color=black]{\textit{mn}}.

As a reference, we shall first describe the spectral anisotropy of the undoped sample [top panels in Figs.~\ref{fig:optan}(a),(b)]. The first positive structure, labelled \fbox{a}, arises from the fundamental optical transition \CCircled[outer color=black]{\textit{11}} [see inset of Fig.~\ref{fig:optan}(a)] which involves the almost purely LH state. This is also an intense transition due to the overlapping envelope functions components (see Fig.~\ref{fig:envbs_kz0}, first row). The positive anisotropy is $\beta \simeq 3/5$, which is expected from the ratio between the momentum matrix element in the $z$ and $x$ directions,
\begin{equation*}
    \bigg\lvert\matrixel{S,\pm \frac{1}{2}}{p_z}{\frac{3}{2}, \pm \frac{1}{2}}\bigg\rvert^2 = 4 \bigg\lvert\matrixel{S,\pm \frac{1}{2}}{p_x}{\frac{3}{2}, \pm \frac{1}{2}}\bigg\rvert^2 \,,
\end{equation*}
hence $\beta=\frac{4-1}{4+1}$.

The next two negative dips in the anisotropy structure \fbox{a} involve the $m=1$ EL subband, and arise from the HH components of transitions \CCircled[outer color=black]{\textit{13}} and \CCircled[outer color=black]{\textit{16}} (see Fig.~\ref{fig:envbs_kz0}, third and sixth row). As HH components  do not couple to EL states for light linearly polarized along $z$, we indeed expect the anisotropy to be negative for these optical transitions.

A second, positive anisotropy set of structures at higher photon energies, labelled \fbox{b}, involves transitions to the $m=2$ conduction subband with  predominantly LH initial states, namely \CCircled[outer color=black]{\textit{22}} and \CCircled[outer color=black]{\textit{24}} transitions (see also Fig.~\ref{fig:envbs_kz0}, second and fourth row). 

\begin{figure}[htp]
in\includegraphics[scale=0.55]{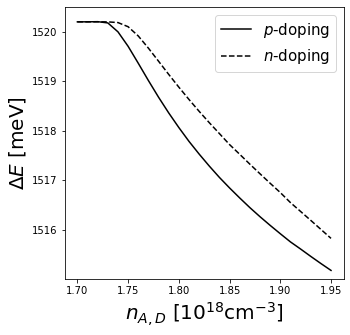}
\caption{\label{fig:Egap} Effective energy gap $\Delta E$ as a function of doping concentration for $n$- and $p$-doped samples. }
\label{fig:energy gap}
\end{figure}

As the optical anisotropy discriminates specific transitions, it is interesting to discuss how the anisotropy spectra evolves with doping concentration. As seen in Fig.~\ref{fig:optan}(a,b) in both $n$- and $p$-doped samples the absorption edge  experiences a red-shift with increasing doping, due to band gap renormalization. 
In Fig.~\ref{fig:energy gap} we compare the energy difference $\Delta E$ between the ground state energy of the conduction and the valence band, respectively, showing that the effective energy gap decreases almost linearly for both kind of samples in the range of a few meV as doping concentration rises.

In $n$-doped samples, Fig.~\ref{fig:optan}(a), the absorption intensity of the lowest transitions gradually vanishes with doping, which is due to two concomitant effects, i) band-filling due to electron subbands falling below the Fermi level, which inhibits inter-band absorption to these levels, and ii) optical matrix element reduction, which is due to Coulomb repulsion: the free charge distribution in the occupied band tends to localize towards the core-shell interfaces as doping concentration is increased, while confining states in the center in the other band, lessening the optical matrix element between initial and final states [Eq.~(\ref{eq:optmel})]. Both effects contribute to suppress low-energy absorption at high-doping, finally moving the absorption edge to the strongly anisotropic structure \fbox{c}, originated by transitions to the $m=7$ EL subband, namely \CCircled[outer color=black]{\textit{71}}, mainly LH with positive anisotropy, and \CCircled[outer color=black]{\textit{73}}, mainly HH, hence with negative anisotropy.

For $p$-doped samples, see Fig.~\ref{fig:optan}(b), the band-filling effect is less pronounced within the examined range of doping. In fact, even at the highest acceptor density shown in Fig.~\ref{fig:optan}(b), the highest valence subband does not cross the Fermi level (see Fig.~\ref{fig:dopP_levels_chlin}). Here, the suppression of the absorption intensity with positive anisotropy at \fbox{a} is mainly due to reduction of the initial and the final states' overlap, due to an increasing localization towards the core-shell interfaces of the hole ground state envelope function (see Fig.~\ref{fig:dopP_env_kz0}, first row). The first negative dip gradually disappears because the third valence subband loses its HH character with increasing doping (see Fig.~\ref{fig:dopP_env_kz0}, third row and Fig.~\ref{fig:dopP_vale}, second column). The opposite occurs for the second negative anisotropy peak, which persists at high doping, due to the increasing HH character of the sixth hole subband with doping, as already pointed out in Sec.~\ref{sec:Pdopbs}, which in turn increases the optical matrix element for $x$-polarized light.

\section{Conclusions}

We have thoroughly investigated the band structure of doped GaAs-based CSNWs, with an emphasis on the evolution of spin-orbit coupled valence band states with doping, either of $n$- or $p$-type. This is an important piece of information for the characterization of such materials, where doping is still an issue.

Our calculations, performed with a state-of-the-art   Burt-Foreman 8-band \kp description, treat many-body effects at the mean-field level, and extend previous investigations to realistic descriptions of doped materials. The use of a flexible FEM approach, which allows to use non-uniform grids, proved to be numerically efficient at different doping levels. This is clearly an advantage in view of multi-parameter optimization, e.g., by stochastic methods \cite{Goldoni2000,Goldoni2001,ram2006wavefunction}.

In particular, we have investigated a proto-typical CSNW with remote doping. As in corresponding planar heterojunctions, the conduction subbands feature a parabolic in-wire dispersion, while hole subbands have a complex dispersion, with inverted masses, which has been rationalized in terms of HH-LH mixing. In large core nanowires, with small confinement energies, increasing doping density moves the majority carriers to the core-shell interface in order to reduce the Coulomb energy.  Correspondingly, the states of the minority carrier band are confined to the core by the self-consistent electrostatic field, and in general the overlap of conduction and valence states decreases. While this is qualitatively true for both types of dopings, our calculations allow to identify several differences between the two type of samples which may have an impact, in particular, on optical absorption. In particular, for $p$-doping the valence band structure is strongly reshaped by confinement of holes at the core-shell interface, and all low energy excitations have a strong LH character.

It may be expected that band structure affects optical absorption and, in particular, optical anisotropy for light polarized along or normal to the nanowire axis. Hence, we have evaluated the doping-density dependent optical anisotropy, which is able to distinguish the spin-orbital character of the transition. In addition to the expected band-filling effects, specific signatures can be identified in the anisotropy patterns which distinguish between $n$- and $p$-doping.

\begin{acknowledgements}

Discussions with Pawe{\l} W\'ojcik are gratefully acknowledged. AB acknowledges partial financial support from the EU project IQubits (Call No. H2020-FETOPEN-2018-2019-2020-01, Project ID No. 829005). The authors acknowledge CINECA for HPC computing resources and support under the ISCRA initiative (No. IsC87 ESQUDO–HP10CXQWD5). 

\end{acknowledgements}

\appendix

\section{\label{app:A} \kp Hamiltonian}
In this appendix we show how to obtain the operator $\hat{H}_{BF}$ in Eq.~(\ref{eq:multibeq}).
In the cartesian basis
\begin{equation}
    \lambda = \left\lbrace \ket{S\!\uparrow}, \ket{S\!\downarrow},  \ket{X\!\uparrow}, \ket{Y\!\uparrow}, \ket{Z\!\uparrow}, \ket{X\!\downarrow}, \ket{Y\!\downarrow}, \ket{Z\!\downarrow} \right\rbrace
    \label{eq:cartesian}
\end{equation}
the Burt-Foreman Hamiltonian with the principal axis along the $[001]$ direction can be written as
\begin{equation}
H^\lambda = H_{0} + V + H_{so}\,.
\label{eq:hamtot}
\end{equation}
The \kp Hamiltonian $H_{0}$, neglecting bulk inversion asymmetry terms, is given by \cite{birner2014multi}
\begin{equation}
H_{0} =
    \begin{pmatrix}
         \kvecT A_{c} \kvec  & 0 & i P \kvecT & \zerovecT \\
        0 & \kvec^T A_{c} \kvec & \zerovecT & i P \kvecT \\
        -i \kvec P & \zerovec & H_v & \zerovec_{3\cross3} \\
        \zerovec & -i \kvec P & \zerovec_{3\cross3} & H_v 
    \end{pmatrix} \\ \,,
\label{eq:h0}
\end{equation}
where $\mathbf{k} = (k_x, k_y, k_z)^\text{T}$, $P$ is the optical matrix parameter, related the the Kane energy parameter by
\begin{equation}
    P = \sqrt{\frac{\hbar}{2 m_0} E_p}
\end{equation}
and $A_c$ is the renormalized conduction band effective mass parameter,
\begin{equation}
    A_c = \frac{\hbar^2}{2 m_e} -\frac{2}{3} \frac{P^2}{E_g} -\frac{1}{3} \frac{P^2}{E_g + \Delta_{so}}\,.
\end{equation}
For many relevant semiconductors, including the present case, the standard parameters lead to a negative value for $A_c$. This fact induces spurious solutions \cite{Eissfeller2011} that bend within the band gap for large wave vectors. To avoid these unphysical results we set $A_c = 1$ and rescale the $E_p$ parameter in order to still get the correct conduction band dispersion: \cite{foreman1997elimination, Veprek2007}
\begin{equation}
    E_p^{\text{rsc}} = \frac{E_g (E_g + \Delta_{so})}{E_g + \frac{2}{3} \Delta_{so}} \left( \frac{1}{m_e}-2 \right)\,.
    \label{eq:ep}
\end{equation}
Foreman rigorously showed that this approach is not an approximation and is equivalent to a change of the Bloch basis \cite{Foreman2007}. We checked that for the regimes investigated and with the relatively coarse grids permitted by the use of FEM, no highly oscillatory, discretization-related spurious solutions appears in our calculations \cite{Eissfeller2011}.

In the above expression the matrix $H_v$ reads
\begin{widetext}
\begin{equation}
\begin{gathered}
H_v = \frac{\hbar^2 \mathbf{k}^2}{2m_{0}}  \mathbb{I}_{3\cross3} + 
    \begin{pmatrix}
        k_x L k_x + k_y M k_y + k_z M k_z &  k_x N^+ k_y + k_y N^- k_x &  k_x N^+ k_z + k_z N^- k_x \\
        k_y N^+ k_x + k_x N^- k_y &  k_x M k_x + k_y L k_y + k_z M k_z &  k_y N^+ k_z + k_z N^- k_y \\
        k_z N^+ k_x + k_x N^- k_z &  k_z N^+ k_y + k_y N^- k_z &  k_x M k_x + k_y M k_y + k_z L k_z
    \end{pmatrix}\,. 
\end{gathered}
\end{equation}
\label{eq:A_four}
\end{widetext}
where $L$, $M$, $N^+$ and $N^-$ are the Dresselhaus-Kip-Kittel parameters which read
\begin{equation}
\begin{split}
 L & =\frac{\hbar^2}{2m_{0}}(- \Tilde{\gamma}_1 - 4\Tilde{\gamma}_2 -1)\,, \\
 M & = \frac{\hbar^2}{2m_{0}}(2 \Tilde{\gamma}_2 - \Tilde{\gamma}_1 -1)\,,  \\
 N^+ & = \frac{\hbar^2}{2m_{0}}(-6 \Tilde{\gamma}_3 - (2 \Tilde{\gamma}_2 - \Tilde{\gamma}_1 -1))  \,,  \\
  N^-& = \frac{\hbar^2}{2m_{0}}(2 \Tilde{\gamma}_2 - \Tilde{\gamma}_1 -1)\,.
\end{split}
\label{eq:lmnparam}
\end{equation}

Here, the modified Luttinger parameters $\Tilde{\gamma}_i$ are
\begin{equation}
\begin{split}
 \Tilde{\gamma}_1 & = \gamma_1 - \frac{E_p^{\text{rsc}}}{3 E_g}\,, \\
 \Tilde{\gamma}_2 & = \gamma_2 - \frac{E_p^{\text{rsc}}}{6 E_g}\,,  \\
 \Tilde{\gamma}_3 & = \gamma_3 - \frac{E_p^{\text{rsc}}}{6 E_g}\,,
\end{split}
\label{eq:luparam}
\end{equation}
where $E_g$ is the bulk band gap and $E_p^{\text{rsc}}$ the rescaled Kane energy.

In Eq.~(\ref{eq:hamtot}) the last two terms represent respectively the in-plane potential profile due to different band edges of adjacent layer  materials
\begin{equation}
    V = diag[E_c, E_c, \Bar{E}_v, \Bar{E}_v, \Bar{E}_v, \Bar{E}_v, \Bar{E}_v, \Bar{E}_v]\,,
\end{equation}
with $\Bar{E}_v=E_v - \frac{\Delta_{so}}{3}$, and the spin-orbit interaction Hamiltonian:
\begin{equation}
H_{so} = \frac{\Delta_{so}}{3}
    \begin{pmatrix}
        0 & 0 & 0 & 0 & 0 & 0 & 0 & 0 \\
        0 & 0 & 0 & 0 & 0 & 0 & 0 & 0 \\
        0 & 0 & 0 & -i & 0 & 0 & 0 & 1 \\
        0 & 0 & i & 0 & 0 & 0 & 0 & -i \\
        0 & 0 & 0 & 0 & 0 & -1 & i & 0 \\
        0 & 0 & 0 & 0 & -1 & 0 & i & 0 \\
        0 & 0 & 0 & 0 & -i & -i & 0 & 0 \\
        0 & 0 & 1 & i & 0 & 0 & 0 & 0 
    \end{pmatrix}\,. \\
    \label{eq:hso}
\end{equation}

The Hamiltonian $H^{\lambda}$ can be rewritten in the form 
\begin{equation}
\begin{aligned}
H^{\lambda} = \sum_{\alpha, \beta = x,y,z} k_{\alpha} D^{\alpha \beta} k_{\beta} + \sum_{\alpha=x,y,z} F^{\alpha}_{L} k_\alpha + k_{\alpha} F^{\alpha}_{R} + G\,,
\label{eq:htot_collected}
\end{aligned}
\end{equation}
where $D^{\alpha, \beta}$, $F^{\alpha}_{L(R)}$ and $G$ are $ 8 \cross 8$ matrices that can be directly  obtained from Eq.~(\ref{eq:hamtot}) by properly collecting terms involving the same powers of the wave vector's components. In particular, $D^{\alpha, \beta} \neq D^{\beta, \alpha}$ and $F^{\alpha}_{L} \neq F^{\alpha}_{R}$. It should be also noted that these matrices are not Hermitian. Nevertheless, the sums $D^{\alpha, \beta} + D^{\beta, \alpha} $ and $F^{\alpha}_{L} + F^{\alpha}_R$ are indeed Hermitian. 

To treat nanowires oriented along an arbitrary direction we define $\mathbf{k}$ in Eq.~(\ref{eq:htot_collected}) in the rotated coordinate system. We have $\mathbf{r'} = R\, \mathbf{r}$ and $\mathbf{k'} = R\, \mathbf{k}$, where $R$ is the orthogonal rotation matrix 
\begin{equation}
R(\theta, \phi) =
\begin{pmatrix}
\cos{\theta}\cos{\phi} & \sin{\phi}\cos{\theta} & -\sin{\theta}\\
-\sin{\phi} & \cos{\phi} & 0\\
\cos{\phi}\sin{\theta} & \sin{\phi}\sin{\theta} & \cos{\theta}
\end{pmatrix}\,.
\label{eq:A_nine}
\end{equation}
The matrices in Eq.~(\ref{eq:htot_collected}), when expressed in terms of the rotated wave vector, $\mathbf{k'} = R\, \mathbf{k}$, transform according to
\begin{eqnarray}
 D^{\alpha \beta}(\theta, \phi) &=& \sum_{\alpha' \beta'}R_{\alpha \alpha'} D^{\alpha' \beta'} R^{-1}_{\beta' \beta}\,, \\
 F^{\alpha}_{L(R)}(\theta, \phi)&=& \sum_{\alpha'} F^{\alpha'}_{L(R)} R^{-1}_{\alpha' \alpha} \, ,
\end{eqnarray}
where $D^{\alpha \beta}$ and $ F^{\alpha}_{L(R)}$ are the matrices defined in the original coordinate system with principal axis directed along the [001] direction.
For convenience, from now on we will omit the $(\theta, \phi)$ notation, implicitly assuming that each of the matrices $D^{\alpha \beta}$ and $F^{\alpha}_{L(R)}$ is defined in the rotated coordinate system.

The transformation that connects the cartesian basis $\lbrace \lambda \rbrace $ to a new one with the principal axis directed along the $(\theta, \phi)$ direction is given by $\lbrace \gamma \rbrace = AU \lbrace \lambda \rbrace$, 
\begin{equation}
\begin{aligned}
    \lbrace \gamma \rbrace  = \left\lbrace \ket{S'\!\uparrow'}, \ket{S'\!\downarrow'}, \right. &  \ket{X'\!\uparrow'}, \ket{Y'\!\uparrow'}, \ket{Z'\!\uparrow'}, \\ & \left. \ket{X'\!\downarrow'}, \ket{Y'\!\downarrow'},\ket{Z'\!\downarrow'} \right\rbrace \,, \\
\label{eq:cartesian_rotated}
\end{aligned} 
\end{equation}
where $U=diag[1,1,R,R]$ is a standard rotation operator and $A=diag[\Bar{A},\Bar{A} \otimes \mathbb{I}_{3 \cross 3} ]$, where 
\begin{equation}
    \Bar{A} = 
    \begin{pmatrix}
        e^{-i \phi/2}\cos{\theta/2} &         e^{i \phi/2}\sin{\theta/2} \\ 
        -e^{-i \phi/2}\sin{\theta/2} & 
        e^{i \phi/2}\cos{\theta/2}
    \end{pmatrix}\,.
    \label{eq:spin_rot}
\end{equation}
rotates the spin.

We now chose the following symmetry adapted basis that diagonalizes spin-orbit interaction: \cite{Los1996}

\begin{equation}
\begin{aligned}
    \{\chi\} = \Big\lbrace & \elu = \ket{S'\!\uparrow'}, \\
                       & \eld = i \ket{S'\!\downarrow'}, \\
                       & \hhu = \sqrt{1 / 2}\ket{(X'+i Y')\!\uparrow'}, \\
                       & \hhd = i \sqrt{1 / 2}\ket{(X'-i Y')\!\downarrow'}, \\
                       & \lhu = i \sqrt{1 / 6}\ket{(X'+i Y')\!\downarrow' }-i \sqrt{2 / 3}\ket{Z'\!\uparrow'}, \\
                       & \lhhd = \sqrt{1 / 6}\ket{(X'-i Y')\!\uparrow'}+\sqrt{2 / 3}\ket{Z'\!\downarrow'}, \\
                       & \sou = \sqrt{1 / 3}\left(\ket{(X'+i Y')\!\downarrow'}+\ket{Z'\!\uparrow'}\right), \\
                       & \sod = -i \sqrt{1 / 3}\left(\ket{(X'-i Y')\!\uparrow' }-\ket{Z'\!\downarrow'}\right) \Big\rbrace \,.
\end{aligned}
\label{eq:angular_momentum_basis}
\end{equation}

Note that here the total angular momentum is defined with respect to the principal axes in the rotated coordinate system.
It follows that the transformation matrix to go from $\lbrace \gamma \rbrace$ to $\lbrace \chi \rbrace$ is
\begin{equation}
    Q = 
    \begin{pmatrix}
        1 & 0 & 0 & 0 & 0 & 0 & 0 & 0 \\
        0 & i & 0 & 0 & 0 & 0 & 0 & 0 \\
        0 & 0 & i\frac{1}{\sqrt{2}} & \frac{1}{\sqrt{2}} & 0 & 0 & 0 & 0 \\
        0 & 0 & 0 & 0 & 0 &  \frac{1}{\sqrt{2}} & i\frac{1}{\sqrt{2}} & 0 \\
        0 & 0 & 0 & 0 & -i\sqrt{\frac{2}{3}} & i\frac{1}{\sqrt{6}} & -\frac{1}{\sqrt{6}} & 0 \\
        0 & 0 & \frac{1}{\sqrt{6}} & -i\frac{1}{\sqrt{6}} & 0 & 0 & 0 & \sqrt{\frac{2}{3}} \\
        0 & 0 & 0 & 0 & \frac{1}{\sqrt{3}} & \frac{1}{\sqrt{3}} & i\frac{1}{\sqrt{3}} & 0 \\
        0 & 0 & -i\frac{1}{\sqrt{3}} & -\frac{1}{\sqrt{3}} & 0 & 0 & 0 & i\frac{1}{\sqrt{3}} 

    \end{pmatrix}\,.
    \label{eq:angular_momentum_rot}    
\end{equation}

Defining $P=QAU$ to be the transformation matrix from $\lbrace \lambda \rbrace$ to $\lbrace \chi \rbrace$, the matrices in $H^{\lambda}$ expressed in terms of the rotated wave vector $\mathbf{k'}$ transform according to \cite{xu}
\begin{equation}
\begin{split}
D^{\alpha \beta} &\rightarrow  P^* D^{\alpha \beta} P^{T}\,, \\
 F^{\alpha}_{L(R)} &\rightarrow P^* F^{\alpha}_{L(R)} P^{T}\,, \\
 G &\rightarrow  P^* G P^{T}\,.     
\end{split}
\end{equation}

To obtain the operator $\hat{H}_{BF}$ appearing in the envelope function equations we now perform the replacements $(k_{x}\rightarrow -i \pdv{x}, k_{y} \rightarrow -i \pdv{y})$ in Eq.~(\ref{eq:htot_collected}) paying attention to preserve the correct operator ordering. Since $k_z$ is now just a parameter, the Hamiltonian operator after the replacement has the following form 
\begin{equation}
\begin{aligned}
\hat{H}_{BF} = \sum_{\alpha, \beta = x,y} \partial_{\alpha} \Bar{D}^{\alpha \beta} \partial_{\beta} + \sum_{\alpha=x,y} \Bar{F}^{\alpha}_{L} \partial_\alpha + \partial_{\alpha} \Bar{F}^{\alpha}_{R} + \Bar{G}\,,
\label{eq:bfperator}
\end{aligned}
\end{equation}
where 
\begin{equation}
\begin{split}
 \Bar{D}^{\alpha \beta} &= -D^{\alpha \beta}\,, \\
 \Bar{F}^{\alpha}_{L} &= -i ( F^{\alpha}_{L} + k_z D^{z \alpha}) \, ,  \\
 \Bar{F}^{\alpha}_{R} &= -i ( F^{\alpha}_{R} + k_z D^{\alpha z}) \,,  \\
  \Bar{G} &= G + k_z^2 D^{z z} + k_z (F^{z}_{R} + F^{z}_{L})\,.  
\end{split}
\end{equation}

\section{\label{app:B} FEM implementation}

Equations (\ref{eq:multibeq}) and (\ref{eq:poisson}) are solved within the FEM framework \cite{ram2002finite}. Here, one writes the proper action integral $\mathcal{A}$ that generates the above set of coupled differential equations through a variational procedure. For the multi-band \kp equations we have \cite{birner2014multi, ram2006wavefunction}
\begin{equation}
\begin{aligned}
    \mathcal{A} = &\sum_{\mu \nu}  \int  \text{d}\rperp  \psi_{\mu}^{*} \Big[ \sum_{\alpha\beta=x,y} -~\cev{\partial}_{\alpha} \Bar{D}^{\alpha \beta}_{\mu \nu} \vec{\partial}_{\beta}~  \\ 
    & +\sum_{\alpha=x,y} \left( \Bar{F}^{\alpha}_{L, \mu \nu} \vec{\partial}_{\alpha} - \cev{\partial}_{\alpha} \Bar{F}^{\alpha}_{R, \mu \nu} \right) + \Bar{G}_{\mu \nu} - E\, \delta_{\mu \nu} \Big] \psi_{\nu}\, ,
\end{aligned}
\label{eq:kpaction}
\end{equation}
where $\mu$ and $\nu$ indicate the components of the envelope function.
Here, the correct operator ordering is retained if we take the differential operator to act on the left (right) when $\hat{k}_{x,y}$ multiplies $F_{R}^{x,y}$ ($F_{L}^{x,y}$).
It is easy to check that Eq.~(\ref{eq:kpaction}) is equivalent to the original eigenvalue problem Eq.~(\ref{eq:multibeq}) by performing a functional variation of $\mathcal{A}$ with respect to $\psi_{\mu}^{*}$ and invoking the principle of least action. Here, surface terms arising from the integration by parts can be eliminated using the continuity of the envelope function and of the probability current across the interfaces \cite{yi1995finite, ogawa1998finite}. If the wave function is set to zero on the domain boundaries, the boundary surface term vanishes too.

The action integral $\mathcal{A}$ is discretized into $n_{el}$ triangular finite elements of the 2D domain, 
\begin{equation}
    \mathcal{A} =  \sum_{i_{el}}^{n_{el}} \mathcal{A}^{(i_{el})}\,.
    \label{eq:actioniel}
\end{equation}
Within the $i_{el}$ element, each component of the unknown envelope function is approximated using Lagrange linear interpolation polynomials \cite{dhatt2012finite} $N_{j}(\rperp)$ so that 
\begin{equation}
    \psi_{\mu}(\rperp) = \sum_{j=1}^{3} \psi_{\mu j} N_{j}(\rperp)\,,
    \label{eq:psiexpanded}
\end{equation}
where the expansion coefficient $\psi_{\mu j}$ represent the value of the $\mu$-th component of the envelope function at the $j$-th triangle's vertex, also called nodal point.

Using Eqs.~(\ref{eq:kpaction}), (\ref{eq:psiexpanded}), we obtain
\begin{equation}
\begin{aligned}
    \mathcal{A}^{(i_{el})} &= \sum_{\mu \nu} \sum_{i j=1}^{3} \psi_{\mu i}^{*} \left[ \int d\rperp N_{i}(\rperp) \mathcal{L}_{\mu \nu} N_{j}(\rperp) \right] \psi_{\nu j} \\
    &= \sum_{\mu \nu}\sum_{i j=1}^{3} \psi_{\mu i}^{*} \mathcal{M}_{\mu \nu i j}^{(i_{el})} \psi_{\nu j}\,,
\end{aligned}
\label{eq:actionexpanded}
\end{equation}
where $\mathcal{L}_{\mu \nu}$ represents the operators appearing in the integrand of Eq.~(\ref{eq:kpaction}), namely the Lagrangian density.

The total action is given by the sum of each element's contribution. This can be written in a very natural manner in matrix form by imposing inter-element continuity through carefully overlaying the element matrices $\mathcal{M}^{(i_{el})}$ \cite{ram2002finite}. To understand how to construct a global matrix starting from element matrices it is convenient to make a simple example. Let's consider a single component of the envelope function $\psi_{\mu}=\psi$ and two adjacent triangular  elements ($i_{el}=1,2$) having two nodes (and one edge) in common. The action integral for these two elements reads
\begin{equation}
    \mathcal{A} = \sum_{ij=1}^{3} \psi_{i}^{(1)*} \mathcal{M}_{ij}^{(1)} \psi_{j}^{(1)} + \sum_{ij=1}^{3} \psi_{i}^{(2)*} \mathcal{M}_{ij}^{(2)} \psi_{j}^{(2)}\,.
    \label{eq:action_example}
\end{equation}
Since the two elements share two nodes and we require inter-element continuity, we set 
$ \psi_{1}^{(1)} =  \psi_{1}^{(2)}, \psi_{2}^{(1)} =  \psi_{2}^{(2)},$ where it is implicitly assumed that the first and the second node of both elements respectively overlap. 
Using the above conditions, it is possible to rewrite  Eq.~(\ref{eq:action_example}) in the \emph{global} form
\begin{equation}
    \mathcal{A} = \sum_{I J=1}^{4} \psi_{I}^* \mathcal{M}_{I J} \psi_{J}\,,
\end{equation}
where $I$ and $J$ now stand for global node indices and $\mathcal{M}$ is obtained from $\mathcal{M}^{(1)}$ and $\mathcal{M}^{(2)}$ by summing the contributions from the same nodes and collecting the envelope functions on common vertices, e.g., for $I, J = 1,2$ we have $\mathcal{M}_{I J}=\mathcal{M}_{I J}^{(1)}+\mathcal{M}_{I J}^{(2)}$.

From this example it is now easy to see that the action integral in its global form can be written as
\begin{equation}
    \mathcal{A} = \sum_{\mu \nu} \sum_{I J}^{n_{glob}} \psi_{\mu I}^* \mathcal{M}_{\mu \nu I J} \psi_{\nu J}\,,
\end{equation}
where $n_{glob}$ is the total number of nodes on the discretization domain.
We now invoke the principle of stationary action and obtain the equation of motion in algebraic form. We vary the action integral with respect to $\psi_{\mu I}^*$ to obtain simultaneous equation for the coefficients $\psi_{\nu J}$,
\begin{equation}
    \fdv{\mathcal{A}}{\psi_{\mu I}^*} = \sum_{\nu} \sum_{ J}^{n_{glob}} \mathcal{M}_{\mu \nu I J} \psi_{\nu J} = 0 \, .
\end{equation}
Given the particular form of the integrand in Eq.~(\ref{eq:kpaction}), the above expression results in a generalized eigenvalue problem:
\begin{equation}
\sum_{\nu} \sum_{ J}^{n_{glob}} \left[ \mathcal{H}_{\mu \nu I J} \, \psi_{\nu J} - E \, \delta_{\mu\nu} \mathcal{S}_{\mu\nu I J}\right] \psi_{\nu J} = 0\, 
    \label{eq:GEVP}
\end{equation}
Here, $\mathcal{H}_{\mu \nu I J}$ represents the discretized form of the Burt-Foreman operator $\hat{H}_{BF}^{\mu \nu}$ in Eq.~(\ref{eq:multibeq}), while $\mathcal{S}_{\mu \nu I J}$ is an overlap matrix which is present due to the non-orthogonality of the basis functions $N_{j}$. From Eq.~(\ref{eq:GEVP}) it is clear that the dimension of the problem is given by the number of nodes $n_{glob}$ in the simulation domain, times the number of components of the envelope function.

For the Poisson equation the energy functional to be minimized is given by 
\begin{equation}
\begin{aligned}
    \mathcal{A} = \int \text{d}\rperp \frac{1}{2} \big[ \epsilon(\rperp)&\grad V_{el}(\rperp) \cdot \grad V_{el}(\rperp) \\ &-\frac{1}{\epsilon_0} V_{el}(\rperp) \rho(\rperp) \big]\,.
\end{aligned}
\end{equation}
A functional variation of $\mathcal{A}$ with respect to $V_{el}$ followed by an integration by parts gives the Poisson equation Eq.~(\ref{eq:poisson}).
Expressing the electrostatic potential again in terms of Lagrange linear interpolation polynomials inside each triangular element,
\begin{equation}
    V_{el}(\rperp)=\sum_{j=1}^{3} V_{el}^{j} N_{j}(\rperp)\, ,
\end{equation}
and following the same procedure described for the \kp problem, a linear system of $n_{glob}$ equations is obtained:
\begin{equation}
    \sum_{J}^{n_{glob}} C_{I J} V_{el} ^{J} = b_{I}\,.
    \label{eq:linsys}
\end{equation}

After the inclusion of proper boundary conditions, Eqs.~(\ref{eq:GEVP}) and (\ref{eq:linsys}) are finally solved with standard library routines.

\bibliography{apssamp}% Produces the bibliography via BibTeX.

\end{document}